\documentclass[PRX,preprint,amsmath,amssymb,superscriptaddress,longbibliography]{revtex4-1}
\usepackage{graphicx}
\usepackage{bm}
\usepackage{hyperref}
\usepackage{breakurl}
\usepackage{multirow}
\usepackage{enumitem}
\usepackage{verbatim}
\setcitestyle{super} 

\begin{document}

\title{Anisotropic gapping of topological Weyl rings in the charge-density-wave superconductor In$_{x}$TaSe$_{2}$ }

\author{Yupeng Li}
      \thanks{Equal contributions}
      \affiliation{Zhejiang Province Key Laboratory of Quantum Technology and Device, Department of Physics, Zhejiang University, Hangzhou 310027, China}

\author{Yi Wu}
      \thanks{Equal contributions}
      \affiliation{Zhejiang Province Key Laboratory of Quantum Technology and Device, Department of Physics, Zhejiang University, Hangzhou 310027, China}

\author{Chenchao Xu}
      \thanks{Equal contributions}
      \affiliation{Zhejiang Province Key Laboratory of Quantum Technology and Device, Department of Physics, Zhejiang University, Hangzhou 310027, China}

\author{Ningning Liu}
      \affiliation{Key Laboratory of Artificial Structures and Quantum Control (Ministry of Education), Department of Physics and Astronomy, Shanghai Jiao Tong University, Shanghai 200240, China}

\author{Jiang Ma}
      \affiliation{Zhejiang Province Key Laboratory of Quantum Technology and Device, Department of Physics, Zhejiang University, Hangzhou 310027, China}

\author{Baijiang Lv}
      \affiliation{Zhejiang Province Key Laboratory of Quantum Technology and Device, Department of Physics, Zhejiang University, Hangzhou 310027, China}

\author{Gang Yao}
      \affiliation{Key Laboratory of Artificial Structures and Quantum Control (Ministry of Education), Department of Physics and Astronomy, Shanghai Jiao Tong University, Shanghai 200240, China}

\author{Yan Liu}
      \affiliation{Zhejiang Province Key Laboratory of Quantum Technology and Device, Department of Physics, Zhejiang University, Hangzhou 310027, China}

\author{Hua Bai}
      \affiliation{Zhejiang Province Key Laboratory of Quantum Technology and Device, Department of Physics, Zhejiang University, Hangzhou 310027, China}

\author{Xiaohui Yang}
      \affiliation{Zhejiang Province Key Laboratory of Quantum Technology and Device, Department of Physics, Zhejiang University, Hangzhou 310027, China}

\author{Lei Qiao}
      \affiliation{Zhejiang Province Key Laboratory of Quantum Technology and Device, Department of Physics, Zhejiang University, Hangzhou 310027, China}

\author{Miaocong Li}
      \affiliation{Zhejiang Province Key Laboratory of Quantum Technology and Device, Department of Physics, Zhejiang University, Hangzhou 310027, China}

\author{Linjun Li}
      \affiliation{State Key Laboratory of Modern Optical Instrumentation, College of Optical Science and Engineering, Zhejiang University, Hangzhou 310027, China}

\author{Hui Xing}
      \affiliation{Key Laboratory of Artificial Structures and Quantum Control (Ministry of Education), Department of Physics and Astronomy, Shanghai Jiao Tong University, Shanghai 200240, China}

\author{Yaobo Huang}
      \affiliation{Shanghai Institute of Applied Physics, CAS, Shanghai, 201204, China}

\author{Junzhang Ma}
      \affiliation{Paul Scherrer Institute, Swiss Light Source, CH-5232 Villigen PSI, Switzerland}

\author{Ming Shi}
      \affiliation{Paul Scherrer Institute, Swiss Light Source, CH-5232 Villigen PSI, Switzerland}

\author{Chao Cao}
      \email{ccao@hznu.edu.cn}
      \affiliation{Department of Physics, Hangzhou Normal University, Hangzhou 310036, China}

\author{Yang Liu}
      \email{yangliuphys@zju.edu.cn}
      \affiliation{Zhejiang Province Key Laboratory of Quantum Technology and Device, Department of Physics, Zhejiang University, Hangzhou 310027, China}

\author{Canhua Liu}
      \affiliation{Key Laboratory of Artificial Structures and Quantum Control (Ministry of Education), Department of Physics and Astronomy, Shanghai Jiao Tong University, Shanghai 200240, China}
      \affiliation{Collaborative Innovation Centre of Advanced Microstructures, Nanjing University, Nanjing 210093, China}

\author{Jinfeng Jia}
      \affiliation{Key Laboratory of Artificial Structures and Quantum Control (Ministry of Education), Department of Physics and Astronomy, Shanghai Jiao Tong University, Shanghai 200240, China}
      \affiliation{Collaborative Innovation Centre of Advanced Microstructures, Nanjing University, Nanjing 210093, China}

\author{Zhu-An Xu}
      \email{zhuan@zju.edu.cn}
      \affiliation{Zhejiang Province Key Laboratory of Quantum Technology and Device, Department of Physics, Zhejiang University, Hangzhou 310027, China}
      \affiliation{Collaborative Innovation Centre of Advanced Microstructures, Nanjing University, Nanjing 210093, China}

\date{\today}

\begin{abstract}
Topological materials and topological phases have recently become a hot topic in condensed matter physics. In this work, we report a topological nodal-line semimetal
In$_{x}$TaSe$_{2}$, in the presence of both charge density wave (CDW) and superconductivity.
In the $x$ = 0.58 samples, the 2 $\times \sqrt{3}$ commensurate CDW (CCDW) and the $2 \times 2$ CCDW are observed below 116 K and 77 K,
respectively. Consistent with theoretical calculations, the spin-orbital coupling gives rise to two two-fold-degenerate nodal rings (Weyl rings) connected by drumhead surface states, confirmed by angle-resolved photoemission spectroscopy. Our results suggest that the $2 \times 2$ CCDW ordering gaps out one Weyl ring in accordance with the CDW band folding, while the other Weyl ring remains gapless with intact surface states. In addition, superconductivity emerges at 0.91 K, with the upper critical field deviating from the s-wave behavior at low temperature, implying possibly unconventional superconductivity. Therefore, In$_{x}$TaSe$_{2}$ represents an interesting material system to study the interplay between CDW, nontrivial band topology and superconductivity.

\end{abstract}

\maketitle

Since topological nodal-line semimetal (TNLSM) was theoretically
proposed in 2011 \cite{TNLSM_BurkovAA_PRB2011}, experimental
verification of theoretically predicted TNLSMs has been achieved,
such as PbTaSe$_{2}$ \cite{PbTaSe2_BianG_NatC}, ZrSiS
\cite{ZrSiS_SchoopLM_NatC}, PtSn$_{4}$ \cite{PtSn4_WuY_NatP}, etc.
In TNLSMs, band crossing forms a closed loop (nodal line) in the
Brillouin zone, which is protected by certain symmetry, such as
mirror refection symmetry, space inversion symmetry, time-reversal
symmetry \cite{TNLS_FangC_CPB2016}, etc. The nodal line can be
fully gapped or gapped into nodal points (Dirac points
\cite{SrIrO3_Carter_PRB2012} or Weyl points
\cite{WeylSemi_WHM_PRX,TaAsDingH_ARPESWSM,TaAs_Hasan_Science,NbP_PRB_WZ})
if the protecting symmetry is broken, resulting in fascinating
properties such as one-dimensional (1D) Fermi arc
\cite{TaAs_Hasan_Science}, negative magnetoresistance
\cite{NMRinWeyl_SonDT_PRB13,Na3Bi_Xiongjun_Science2015,WeylSemi_WHM_PRX,NMR_YPL_FP17},
etc. While TNLSMs could exhibit drumhead surface states
\cite{TlTaSe2_BianG_PRB16,Ca3P2nodalline_ChanYH_PRB16}, anomalous
quantum oscillations
\cite{TNLSMphase_LiC_PRL2018,TNLSMQO_YangH_PRL2018} and so on,
they could also host topological superconductivity
\cite{PbTaSe2_GuanSY_SciAdv16,Yplisummary_AQT19} and
three-dimensional quantum Hall effect (3D QHE)
\cite{NodalLineSM3DQHE_Molina_PRL18}. The TNLSM PbTaSe$_{2}$
\cite{PbTaSe2_BianG_NatC} is a promising candidate for topological
superconductors (TSCs) because not only the fully gapped
superconductivity is confirmed
\cite{PbTaSe2SC_AMN_PRB14,PbTaSe2NSC_WangMX_PRB16,PbTaSe2Nodeless_PangGM_PRB16},
but also the zero-energy Majorana bound states are detected in the
vortices \cite{PbTaSe2_GuanSY_SciAdv16}. In addition, drumhead
surface states, which are nontrivial surface states connected by
bulk nodal lines
\cite{PbTaSe2_BianG_NatC,TlTaSe2_BianG_PRB16,Ca3P2nodalline_ChanYH_PRB16},
may give rise to 3D QHE when the magnetic field is parallel to the
nodal lines \cite{NodalLineSM3DQHE_Molina_PRL18}.

\begin{figure*}[!thb]
\begin{center}
\includegraphics[width=6in]{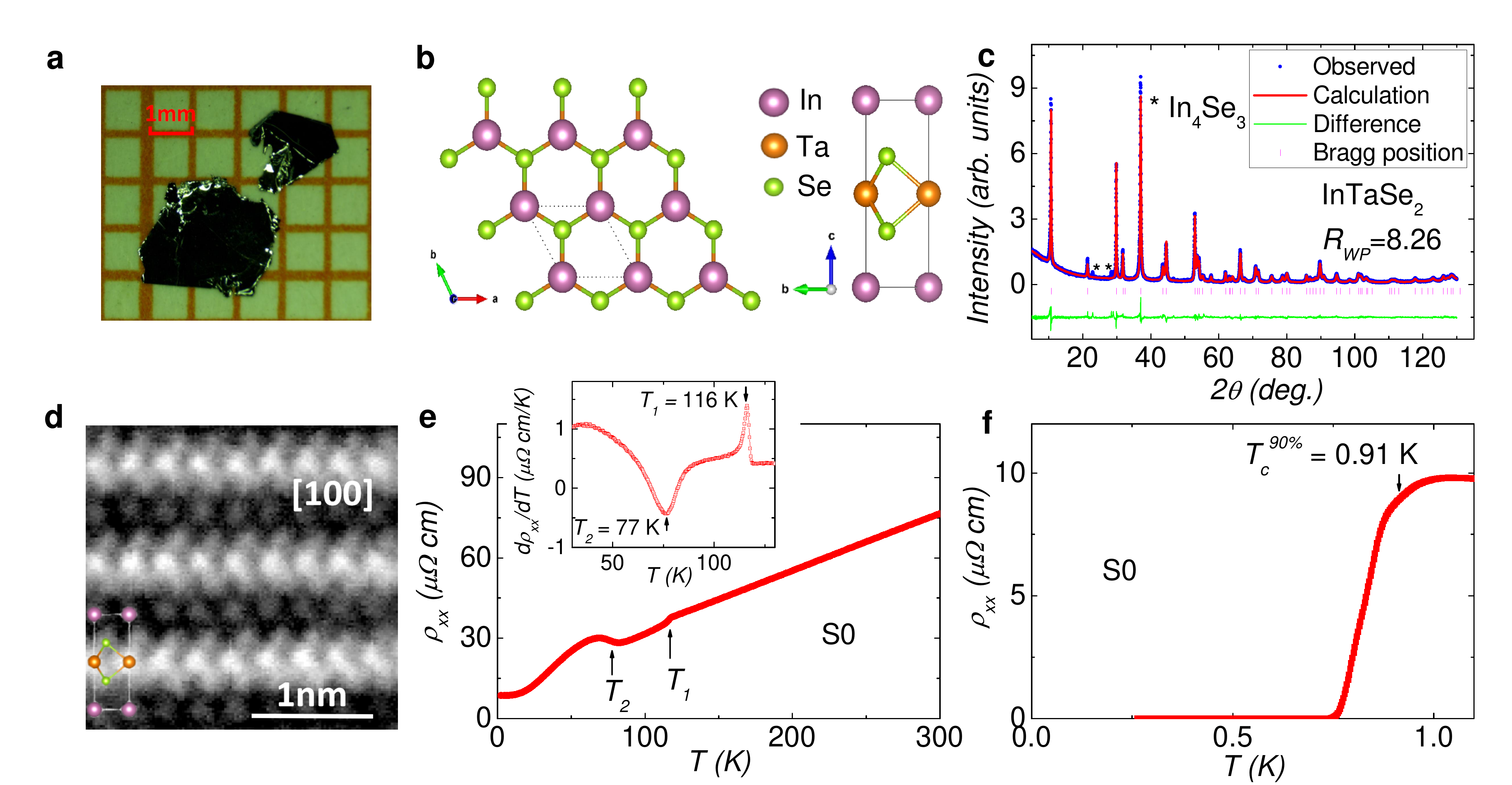}
\end{center}
\caption{\label{Fig1}\textbf{Characterization of In$_{0.58}$TaSe$_{2}$.} \textbf{a}, The optical photograph of $x=0.58$ compounds. \textbf{b}, Top view and side view of the crystalline structure of InTaSe$_{2}$.
\textbf{c}, Good Rietveld refinement of powder XRD data for
polycrystalline InTaSe$_{2}$. \textbf{d}, A HRTEM image of
the In$_{0.58}$TaSe$_{2}$ single crystal along the [100] projection confirms this crystal structure. \textbf{e}, The resistivity
of In$_{0.58}$TaSe$_{2}$ shows two CDW transitions, and $T_{1}$ = 116 K and $T_{2}$ = 77 K are obtained by
the differential resistivity in the inset. \textbf{f}, The superconducting transition emerges at 0.91 K.   }
\end{figure*}

Charge density wave (CDW), being a ubiquitous ordered state in
many condensed matter systems, typically develops in
low-dimensional systems and minimizes the energy by opening up
energy gaps near Fermi level \cite{CDW_GrunerG_RMP1988}. While it
is traditionally believed to compete with superconductivity
\cite{CDWToSc_MorosanE_Nature06,CDWCopSC_ChangJ_NP12} in many
layered structure systems (such as cuprates
\cite{CDWCeoxSC_Howald_PNAS03}, transition-metal chalcogenides
\cite{CDWCeoxSC_KissT_NP07}), they coexist and are very likely
intimately connected. CDW is also theoretically proposed to induce
novel topological states. For example, in a Weyl semimetal,
axionic quasi-particle can be excited when the Weyl nodes are fully
gapped by CDW, and the axion field preserves the dissipationless
transport of Weyl fermion \cite{AxioninWeyl_WangZ_PRB13},
including chiral anomaly. This axionic-CDW is well identified in
Weyl semimetal (TaSe$_{4}$)$_{2}$I \cite{Ta2Se8I_GoothJ_Nature19}.
Moreover, CDW is also an important ingredient for realizing 3D
QHE, which is first observed in a topological material ZrTe$_{5}$
when a CDW gap is induced in the magnetic field direction
\cite{3DQHE_TangF_arXiv18}. Therefore, TNLSMs with CDW states are
worth searching for to detect fantastic topological phases.

In this work, we synthesize a TNLSM In$_{x}$TaSe$_{2}$ hosting both CDW ordering and superconductivity.
Our density functional theory (DFT)
calculations demonstrate that InTaSe$_{2}$ exhibits two separate nodal rings (Weyl rings) in the $k_{x}$-$k_{y}$ plane,
similar to InNbS$_{2}$ \cite{InTaSe2_DuYP_PRB17} and TlTaSe$_{2}$ \cite{TlTaSe2_BianG_PRB16}.
The compound In$_{x}$TaSe$_{2}$ with $x$ = 0.58 exhibits a $2 \times \sqrt{3}$ commensurate CDW (CCDW) state below 116 K accompanied by the
resistance anomaly and the specific-heat jump. With decreasing temperature, a $2 \times 2$ CCDW state with an upturn of resistivity below
approximately 77 K is detected by scanning tunneling microscopy (STM), transport measurements and
angle-resolved photoemission spectroscopy (ARPES) measurements. Our data indicate that the
outer Weyl ring could be anisotropically gapped out due to the $2 \times 2$ CCDW state,
while the inner Weyl ring remains gapless with associated surface states,
consistent with the DFT calculations. Furthermore, the superconducting transition emerges at 0.91 K in In$_{0.58}$TaSe$_{2}$, implying it is a potential candidate for TSCs.

\begin{figure*}[!thb]
\begin{center}
\includegraphics[width=6in]{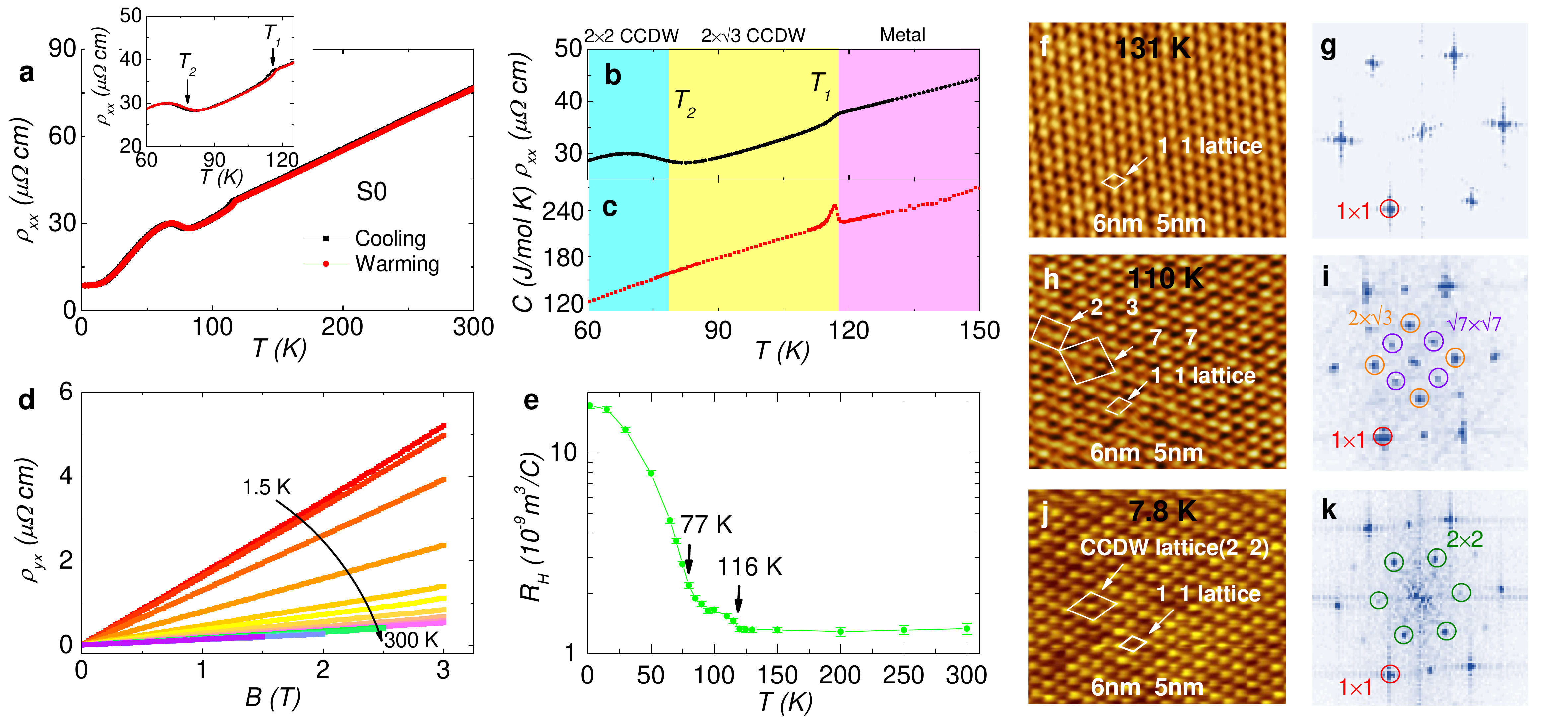}
\end{center}
\caption{\label{Fig2}\textbf{CDW states of In$_{0.58}$TaSe$_{2}$.} \textbf{a}, $\rho_{xx}$ of In$_{0.58}$TaSe$_{2}$ for temperature warming and cooling
at zero magnetic field is measured, and the inset is the enlarged view. Enlarged parts of
$\rho_{xx}$ (\textbf{b}) and specific heat (\textbf{c}) in In$_{0.58}$TaSe$_{2}$ are plotted from 60 to 150 K with a $2 \times 2$ CCDW state, a $2 \times \sqrt{3}$
CCDW state and a metal state. A distinct jump exists at $\sim$ 116 K in the specific heat at zero magnetic field. \textbf{d}, Magnetic-field dependence
of $\rho_{yx}$ at different temperatures. \textbf{e}, Hall coefficient vs temperature increases much faster below 120 K.
\textbf{f},\textbf{h},\textbf{j}, STM images (V = -500mV, I = 100 pA; V = -500mV, I = 100 pA; and V = 600mV, I = 100 pA) of the In$_{0.58}$TaSe$_{2}$ surface at 131 K, 110K and 7.8 K
are plotted, respectively, and their FFT images are \textbf{g}, \textbf{i} and \textbf{k}, respectively. The pattern marked by red circle is
1$\times$1 lattice of In$_{0.58}$TaSe$_{2}$. The orange and purple circles represent the $2 \times \sqrt{3}$ and $\sqrt{7} \times \sqrt{7}$ CCDW pattern in \textbf{i}, respectively.
Green circles are the $2 \times 2$ CCDW pattern in \textbf{k}. }
\end{figure*}

\begin{figure*}[!thb]
\begin{center}
\includegraphics[width=6in]{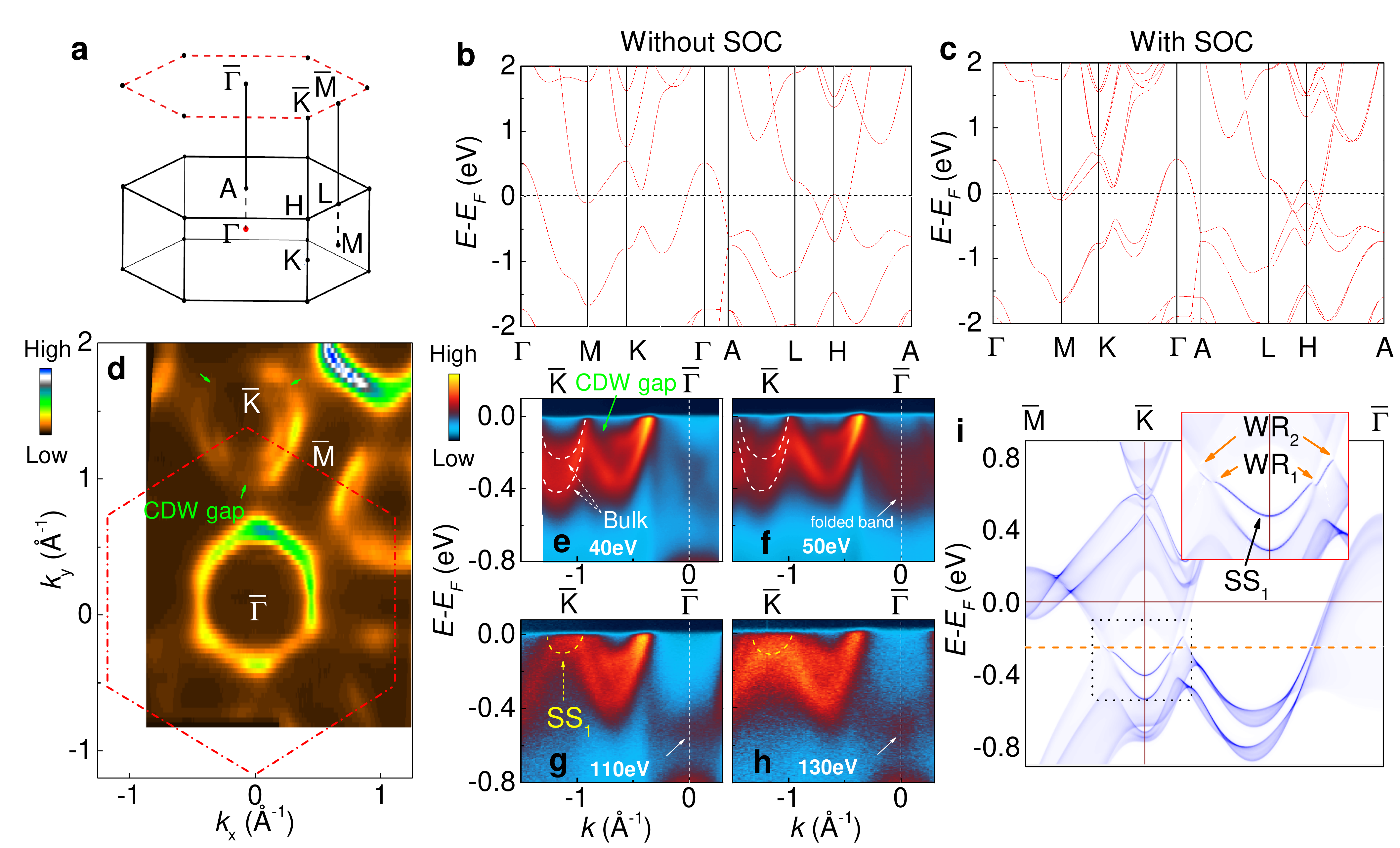}
\end{center}
\caption{\label{Fig3}\textbf{Band structures of In$_{0.58}$TaSe$_{2}$.} \textbf{a}, First Brillouin zone of InTaSe$_{2}$ and the projection of
the (001) surface Brillouin zone. \textbf{b},\textbf{c}, The calculated band structures of InTaSe$_{2}$
are displayed without SOC and with SOC. The Dirac-type nodal ring forms around
$H$ point in \textbf{b}, and SOC splits the Dirac-type nodal rings into two Weyl rings in \textbf{c}.
\textbf{d}, ARPES Fermi surface contour is measured with 65-eV photons, and the CDW gap shows up as suppression of spectral intensity in the Fermi surface, marked by green arrows. \textbf{e},\textbf{f},\textbf{g},\textbf{h}, ARPES energy-momentum cut along $\bar{M}-\bar{K}-\bar{\Gamma}$
line at four representative photon energies. Surface state SS$_{1}$ cound be observed near the Fermi level at 110 eV and 130 eV.
The In 5p bulk band near $\bar{K}$ point is marked by white dash line, and the CDW band gap exists
along $\bar{K}-\bar{\Gamma}$ while the CDW folded band at $\bar{\Gamma}$ is also marked by the white arrow. \textbf{i}, Compared with ARPES spectra, calculated (001) band structures alone $\bar{M}-\bar{K}-\bar{\Gamma}$
reveals that the actual Fermi level (orange dashed line) seems to be lower by $\sim$ 0.25 eV compared to DFT calculations of InTaSe$_{2}$.
Except from the hole doping, the experimental band structure is consistent with the calculation. The inset is the enlarged view of the dotted rectangular, highlighting the SS$_{1}$.   }
\end{figure*}

\vspace{3ex}

\noindent\textbf{Sample characterizations}

\noindent The compound In$_{x}$TaSe$_{2}$
is synthesized with the hexagonal structure
\emph{P$\bar{6}$m}2 (space group no.187), as shown in Fig. 1a.
From the side view of the crystal structure in Fig. 1b, the
noncentrosymmetric structure of In$_{x}$TaSe$_{2}$ shows adjacent
TaSe$_{2}$ layers are intercalated with an In layer, and Ta atoms
align with In atoms in In$_{x}$TaSe$_{2}$, like the structure of
TlTaSe$_{2}$ \cite{TlTaSe2_BianG_PRB16} and InNbS$_{2}$
\cite{InTaSe2_DuYP_PRB17}. In contrast, the Se atoms are aligned
with Pb atoms in PbTaSe$_{2}$ \cite{PbTaSe2SC_AMN_PRB14}. This
structure is clearly identified by the Rietveld refinement of
powder XRD data for polycrystalline InTaSe$_{2}$ in Fig. 1c,
which has a reliable factor $R_{wp} =$ 8.26\%. The refined
structural parameters are $a$ = $b$ = 3.4605 \AA \ and $c$ =
8.3231 \AA. From the EDS data in Supplementary Fig. S1, the
chemical composition In:Ta:Se = 0.58:1:2.04, and a number of
samples grown by the vapor transport method are obtained with $x
\sim 0.6$, as shown in table I of Supplementary Information.
Moreover, a HRTEM image in Fig. 1d confirms this structure in a
In$_{0.58}$TaSe$_{2}$ single crystal sample. We do not find any
sign of mixed phase in both XRD (see Supplementary Fig. S1) and HRTEM data, indicating that the indium atoms are
probably distributed uniformly within the intercalated layer. In
Fig. 1e, resistivity measurements show two interesting CDW-like
transitions in In$_{0.58}$TaSe$_{2}$ at $T_{1} = 116$ K and $T_{2}
= 77$ K, which is acquired by the differential resistivity in the
inset of Fig. 1e, and the further verification of CDW states will
be given in the following subsection. Intriguingly, a
superconducting transition at 0.91 K (90\% of normal state
resistivity) is observed in this topological semimetal, as shown in Fig. 1f.

\vspace{3ex}

\noindent\textbf{Signature of CDW states}

\noindent We now present the study of the CDW
transitions in our samples. In Fig. 2a, the resistivity of
In$_{0.58}$TaSe$_{2}$ exhibits two transitions below 120 K. The
inset of Fig. 2a shows a very weak hysteresis at $T_{1}$ and the
hysteresis is absent at $T_{2}$ between temperature warming and
cooling measurements. According to the temperature warming data,
there is a decrease in resistivity at $T_{1}$ of 116 K and an
upturn at $T_{2}$ of 77 K. The transition at $T_{1}$ characterized
by a $\lambda$-like jump is also observed in the specific heat
shown in Fig. 2c. The similar transitions are revealed in
2H-TaSe$_{2}$, where an incommensurate CDW (ICCDW) state emerges
at approximately 121 K, and a CCDW transition
appears at approximately 90 K accompanied by a slight kink on
resistivity \cite{TaSe2series_Harper_PRB77,TaSe2-xSx_LiL_njp17}.
Therefore, the transition features in
In$_{0.58}$TaSe$_{2}$ are CDW-like transitions but a little
different from 2H-TaSe$_{2}$. From the positive Hall
resistivity of In$_{0.58}$TaSe$_{2}$ at various temperatures in
Fig. 2d, the hole-type charge carrier is dominant in this
system. The Hall coefficient $R_{H}$ of In$_{0.58}$TaSe$_{2}$
begins to increase below $T_{1}$ and then increases much faster
below $T_{2}$ in Fig. 2e, suggesting opening of the CDW gap.

Subsequently, STM measurements are performed to detect the CDW
states of In$_{0.58}$TaSe$_{2}$ with atomic resolution topographs
and their fast Fourier transforms (FFTs) at different temperatures,
as shown in Fig. 2f-k. In Fig. 2f, a good morphology feature
is obtained at 131 K with a clear 1$\times$1 atomic arrangement of
this compound. This arrangement is marked by a red circle in Fig.
2g, the FFT of Fig. 2f. The wave vector of the red circle is
$\vec{Q}_{0}=4\pi/\sqrt{3}a$, where $a$ is the lattice constant.
At 110 K, there are two types of CDW states in Fig. 2h, which
represent a $2 \times \sqrt{3}$ (orange circles) and a $\sqrt{7}
\times \sqrt{7}$ (purple circles) CCDW pattern marked in Fig. 2i
according to further analysis. The $2 \times \sqrt{3}$ CCDW state
is likely more intrinsic, because it can be transformed to the
$\sqrt{7} \times \sqrt{7}$ pattern in some domains where the
diagonal position signals of $2 \times \sqrt{3}$ CCDW lattices
become weaker \cite{CDWInTaSe2}. The similar CDW state is
also observed in the Li-doped 2H-TaSe$_{2}$
\cite{2HTaSe2Lidoping_HallJ_ACSNano19}. When the temperature is lowered to
7.8 K, a 2$\times$2 CCDW superlattice develops in the
$k_{x}-k_{y}$ plane in Fig. 2j, whose FFT pattern in Fig. 2k
exhibits this CCDW pattern (green circles) with a wave vector
(0.5, 0, 0) \cite{NbSe2STM_SA_PNAS13}. These CDW states could not
be well explained by Fermi surface nesting, as illustrated in
Supplementary Fig. S2, where no distinct CDW-type peak is observed
in the calculated imaginary part of susceptibility. Thus the CDW
mechanism of this compound needs further investigations
\cite{CDWorigin_Rossnagel_JPCM2011,TaSe2nesting_Mazin_PRB08}.

\vspace{3ex}

\noindent\textbf{DFT calculations and ARPES measurements}

\noindent The calculated bulk band structures of InTaSe$_{2}$
without and with  spin-orbital coupling (SOC) are shown in Fig. 3b and Fig. 3c, respectively.
In Fig. 3b, a hole pocket and a electron pocket are formed
at $H$ point by Ta-5d orbitals and In-5p orbitals, respectively.
Subsequently, the hybridization of these two bands results
in the band inversion and a Dirac-like nodal
ring in the $k_{z}$ = $\pi$ plane. The similar band structures at $H$ point
are also observed in PbTaSe$_{2}$ \cite{PbTaSe2_BianG_NatC}.
When SOC is included in Fig. 3c, this Dirac-like nodal ring will split into two
two-fold-degenerate nodal rings, which are also named Weyl rings,
hosting a Berry phase of $\pi$ for each Weyl ring \cite{InTaSe2_DuYP_PRB17,TlTaSe2_BianG_PRB16,Ca3P2nodalline_ChanYH_PRB16}.
These rings are protected against band gap opening by the mirror symmetry \cite{PbTaSe2_BianG_NatC}.

\begin{figure*}[!thb]
\begin{center}
\includegraphics[width=6in]{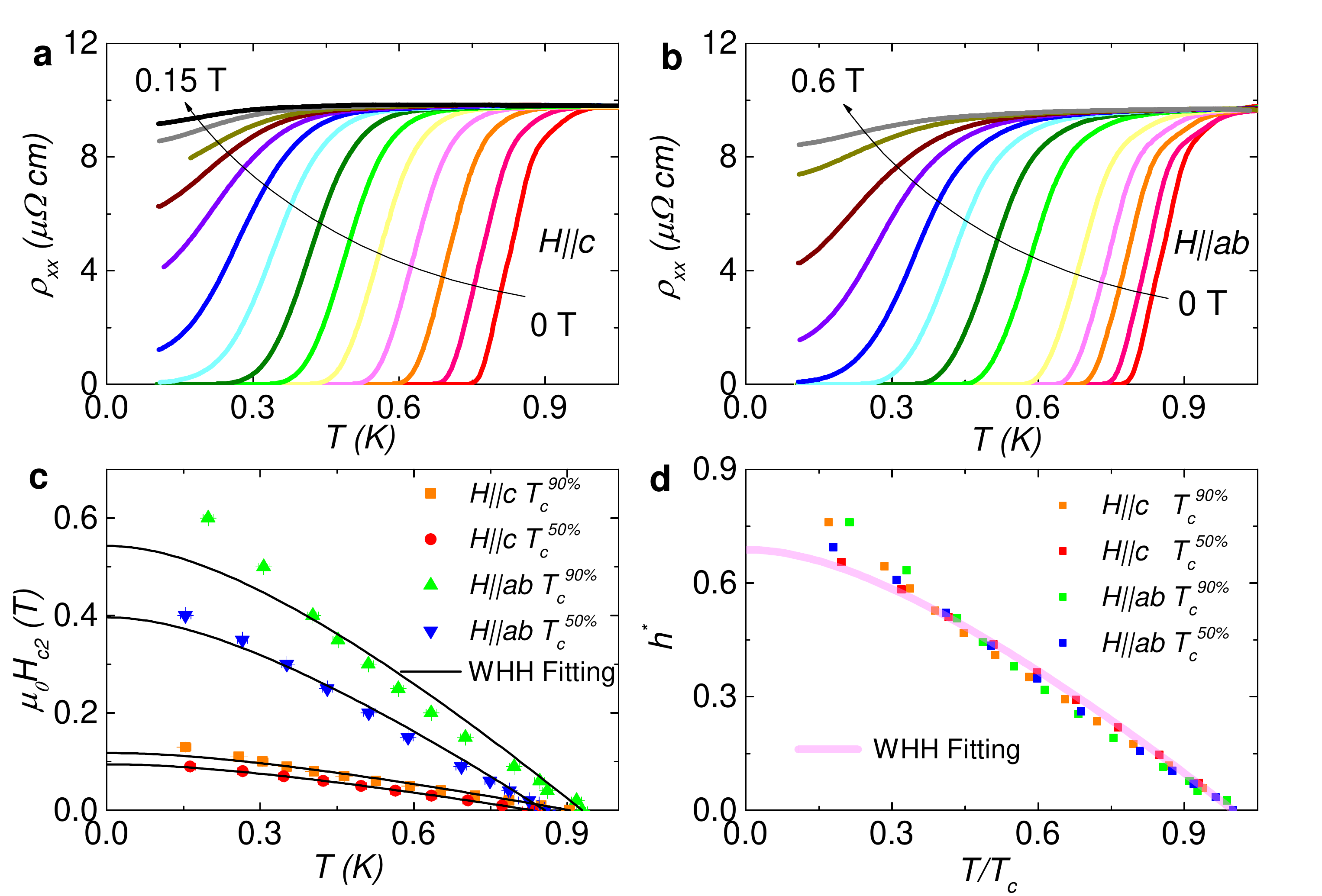}
\end{center}
\caption{\textbf{Superconductivity of In$_{0.58}$TaSe$_{2}$.} \textbf{a},\textbf{b}, The superconducting transitions are measured under different magnetic fields
when $H$ is perpendicular and parallel to $ab$ plane of In$_{0.58}$TaSe$_{2}$,
respectively. \textbf{c}, The upper critical fields are plotted as a function of temperature,
and the WHH formula is used to fit $H_{c2}$ data. The superconducting
transition temperature ($T_{c}$) is extracted at 90\% and 50\% of the normal-state resistivity.
\textbf{d}, Normalized upper critical field as a function of temperature $T/T_{c} $ is plotted for $T_{c}^{90\%}$ and $T_{c}^{50\%}$ at $H \| c$ and $H \| ab$,
and the WHH model is used to fit the data.   }
\label{Fig4}
\end{figure*}

Furthermore, the ARPES measurements are implemented to probe the aforementioned band structures in In$_{0.58}$TaSe$_{2}$.
Fig. 3d displays ARPES spectral intensity in the $k_{x}-k_{y}$ plane at the photon energy of 65 eV and the measured temperature is 20 K. A hole pocket (from Ta 5d bands) and two electron pockets (from In 5p bands) are observed near $\bar{\Gamma}$ and $\bar{K}$, respectively. Band structures along $\bar{M}-\bar{K}-\bar{\Gamma}$ taken at some representative photon energies are summarized in Fig. 3e-h (more systematic ARPES data under different photon energies are presented in Supplementary Fig. S5). Compared with experimental bands in Fig. 3e-h, the calculated fermi energy is a little higher by approximately 0.25 eV in Fig. 3i because In$_{0.58}$TaSe$_{2}$ can be viewed as the hole-doped InTaSe$_{2}$. The other experimental band characteristics are almost consistent with calculated ones after shifting the Fermi energy to approximately $-$0.25 eV (the orange dash line). Specifically, in Fig. 3g and 3h, a band with bottom at $\sim$-0.1 eV (yellow dashed line) can be observed in the 110 eV and 130 eV data, seemingly in agreement with the calculated position of the surface state SS$_{1}$, i.e., the drumhead surface state connecting the bulk Weyl ring (WR$_{1}$, Fig. 3i). According to DFT calculations in the absence of CDW order, two sets of WRs can be expected due to two sets of spin-orbit split bulk bands (Fig. 3c), and these bulk bands are indeed observed experimentally (white dashed lines in Fig. 3e and 3f). The assignment of bulk and surface states is made through detailed photon-energy dependent study and careful comparison with DFT calculations, as shown in Supplementary Fig. S5. Without the CDW order, WR$_{2}$ could also give rise to nontrivial surface states (see Supplementary Fig. S4), which is buried within the bulk band and hence mixes strongly with it. Upon entering the CDW phase, the CDW band folding occurs and gives rise to CDW band gaps in characteristic energy-momentum positions (see Supplementary Fig. S3 for a detailed schematic for CDW band folding). The CDW band folding can be best seen near the $\bar{\Gamma}$ point, where the original band at $\bar{M}$ point is folded due to 2$\times$2 CDW order. The CDW gap is also manifested by the suppressed spectral intensity near WR$_{2}$ (see green arrow in Fig. 3d and 3e), in accord with the 2$\times$2 CDW order (see Supplementary Fig. S3). This gap can be further supported by the temperature-dependent ARPES measurements shown in Supplementary Fig. S6. Due to the observed large CDW gap ($\sim$ 100 meV), the outer Weyl ring appears to be gapped out anisotropically, as shown in Fig. 3d and 3e. However, the inner Weyl ring is far away from the gap region and therefore probably remains gapless with the SS$_{1}$ intact. Since the CDW mostly involves the in-plane atomic movements, which likely preserves the mirror symmetry with respect to the Ta plane, the inner Weyl ring and associated SS can be robust against CDW order \cite{PbTaSe2_BianG_NatC}. Our results therefore demonstrate that under the influence of CDW, the topological nodal rings can be significantly modified and even gapped out anisotropically in the momentum space, in compliance with the CDW order.

\vspace{3ex}

\noindent\textbf{Superconductivity}

\noindent As shown in Fig. 1f,
superconductivity is observed in In$_{0.58}$TaSe$_{2}$ at
$T_{c}$ = 0.91 K. Fig. 4a and 4b show the superconducting
transitions at various magnetic fields ($H$) with $H \| c$ plane
and $H \| ab$ plane, respectively. Fantastically,
superconductivity is more robust with increasing $H$ for $H \| ab$
plane whether $T_{c}$ is extracted by 90\% or 50\% drop of normal
resistivity in Fig. 4c. These upper critical fields
$H_{c2}$ as a function of temperature in Fig. 4c can be well
fitted by the s-wave superconductivity
(Werthammer-Helfand-Hohenberg (WHH) model \cite{WHH_WNR_PR}), and
$H_{c2}(0) \approx 0.7T_{c}dH_{c2}/dT|_{T = T_{c}}$. From these
fittings, the extrapolated zero-temperature $H_{c2}$ are estimated
as $H^{H\parallel c}_{c2}$ ($T_{c}^{90\%}$) = 0.1176 T and
$H^{H\parallel ab}_{c2}$ ($T_{c}^{90\%}$) = 0.5428 T, both of
which are nearly two orders of magnitude larger than 2H-TaSe$_{2}$
\cite{TaSe2_Yokota_2000}. It is interesting that the intercalation
of In atom layer into the 2H-TaSe$_{2}$ results in such enhanced
$H_{c2}$. Moreover, the anisotropy factor $H^{H\parallel
ab}_{c2}$/$H^{H\parallel c}_{c2} = 4.6$ in the sample S0 is larger
than that ($\approx$ 3) in 2H-TaSe$_{2}$ \cite{TaSe2_Yokota_2000},
and this large anisotropic behavior is also observed in other
samples S1 and S2 (see Fig. S7 and S8 of Supplementary
Information). It is a common feature for such layered structure
compounds to host anisotropic superconductivity
\cite{TaSe2_Yokota_2000,TaSe2_BH_JPCM18} or even quasi-2D nature
of superconductivity \cite{Iontronics_bisri_AM2017}. In addition,
a universal function $h^{*}(t)$ can be used to describe all the
$H_{c2}$ data well in Fig. 4d, where $h^{*}(t)=
(H_{c2}/T_{c})/|dH_{c2}/dT|_{T_{c}}$ and $t=T/T_{c}$
\cite{WHH_WNR_PR}, and $h^{*}(0)$ should be equal to $\sim$ 0.7
($\sim$ 0.8) for s-wave superconductor \cite{WHH_WNR_PR} (p-wave
superconductor \cite{PwaveFitting_PR1980}). The large deviations
of $H_{c2}$ from the WHH fitting at low temperatures and
$h^{*}(0)$ close to $\sim$ 0.8 can be observed in Fig. 4d, and
similar behaviors are also discovered in PbTaSe$_{2}$
\cite{PbTaSe2SCP_ZhnagCL_PRB16}, Cu$_{x}$Bi$_{2}$Se$_{3}$
\cite{CuxBi2Se3HP_BayTV_PRL12}, YPtBi \cite{YPtBiSC_BayTV_PRB12},
RPdBi \cite{RPdBi_SciAdv15}, UTe$_{2}$ \cite{UTe2_RanS_Science19}
and so on, which suggests the possible unconventional
superconductivity or spin-triplet superconductivity in In$_{0.58}$TaSe$_{2}$.

\vspace{3ex}

\noindent\textbf{Discussions}

\noindent By combining magneto-transport, STM and ARPES measurements, and the DFT calculations, the coexistence of CDW, nodal-line topological states and superconductivity can be confirmed in In$_{x}$TaSe$_{2}$. While the interplay between nontrivial band topology and superconductivity is currently attracting tremendous research interest, the influence from CDW ordering is much less studied experimentally, despite considerable theoretical interests \cite{AxioninWeyl_WangZ_PRB13}. At high temperature, where the system has not entered the superconducting state, the $2 \times \sqrt{3}$ CCDW and the $2 \times 2$ CCDW are detected by STM below 116 K and 77 K, respectively.
The comparison between ARPES and DFT data shows the existence of Weyl rings and drumhead surface states, confirming the topological nature of its electronic structure. The CDW ordering could cause pronounced anisotropic gapping of the Weyl rings in the momentum space. Such an anisotropically gapped nodal-ring semimetal is a direct manifestation of the interplay between electronic topology and CDW order.
This type of CDW-related topological state offers a new platform to search for fantastic topological states, such as 3D QHE \cite{NodalLineSM3DQHE_Molina_PRL18,3DQHE_TangF_arXiv18}, axionic quasi-particle \cite{Ta2Se8I_GoothJ_Nature19,AxioninWeyl_WangZ_PRB13}, etc.

Below $\sim$ 0.9 K, In$_{0.58}$TaSe$_{2}$ enters the
superconducting state and our upper critical field measurements
exhibit obvious deviation from the WHH fitting at low temperature,
which suggests possibly unconventional superconductivity. Is such
superconducting behavior linked to the topological nodal-line
states, or is it intrinsically a TSC? These
questions remain to be answered in the future. Another important
question is the possible role of CDW in the superconducting state,
and how it could compete or coexist with the superconductivity. It
is prospective that the large and anisotropic CDW gap could have a
profound effect on the (possibly topological) superconductivity in
the system.

\vspace{3ex}

\noindent\textbf{Motheds}

\noindent\textbf{Sample preparation.} The polycrystalline InTaSe$_{2}$ was prepared by the solid-state reaction.
Firstly, the stoichiometric mixture of In (Alfa Aesar 99.95\%), Ta (Alfa Aesar 99.95\%)
and Se (Alfa Aesar 99.95\%) was sealed in a quartz ampoule and then heated to
1123 K for 2 days. Secondly, the resultant was reground and pressed into a pellet,
which was then jacketed by an evacuated ampoule and the polycrystalline sample
was obtained after heating the ampoule at 1123 K for 2 days.
The In$_{0.58}$TaSe$_{2}$ single crystals were synthesized by a vapor transport method.
The polycrystalline InTaSe$_{2}$ was sealed in an evacuated
ampoule with a length of 16 cm using iodine/InBr$_{3}$ as a transport agent (5 mg/cm$^{3}$).
Several platelike single crystals were grown in a two-zone furnace for
three weeks with the powder end at 1123 K and the cooler end at 1023 K.
The samples grown by the transport method have nearly the similar $x \sim 0.6$.

\vspace{3ex}

\noindent\textbf{Measurements.} The structure and composition ratio of samples were analyzed by x-ray diffraction
and an energy-dispersive x-ray spectroscopy (EDS), respectively. High-resolution transmission electron
microscope (HRTEM) measurement were employed at room temperature with an aberration-corrected
FEI-Titan G2 80-200 ChemiSTEM. The longitudinal resistivity and Hall resistivity
were measured using a standard six-probe technique on an Oxford-15T cryostat. The specific heat
was measured for single crystals In$_{0.58}$TaSe$_{2}$ on a physical
properties measurement system (PPMS). STM measurements were conducted in a unisoku-USM1000 STM system.
High-resolution ARPES measurements were carried out on the SIS-HRPES beamline at the Swiss Light Source (SLS),
Paul Scherrer Institute (Switzerland), and the Dreamline at Shanghai Synchrotron Radiation Facility.
The samples were cleaved under ultrahigh vacuum and meassured at low temperatures.
The typical energy and momentum resolution is $\sim$15 meV and $\sim$0.01 \AA$^{-1}$.

\vspace{3ex}

\noindent\textbf{Band calculations.} The DFT calculations were performed using the generalized gradient
approximation (GGA) method under the Perdew-Burke-Ernzerhoff (PBE) parameterization. The band structure results were also checked by modified Becke-Johnson (mBJ) method \cite{mBJ_Tran_PRL102}. A $\Gamma$-centred 18$\times$18$\times$6 Monkhorst-Pack $k$-point mesh was applied in the calculations. During the whole calculation, the lattice constants and the atomic coordinates were from the XRD Rietveld refinements. The surface states were calculated using the surface Green's function \cite{surfacestate_1985JPF}.

\vspace{3ex}

\noindent\textbf{Data availability}

\noindent The data that support the findings of this study are available
from the corresponding author upon reasonable request.

\vspace{3ex}

\noindent\textbf{Acknowledgments}

\noindent We thank Lu Li, Guang Bian, Kunliang Bu, Youting Song
for insightful discussions. This work was supported by the
National Key R\&D Program of the China (Grant No. 2016YFA0300402,
2014CB648400 and 2016YFA0300203), the National Science Foundation
of China (Grant Nos. 11774305 and 11274006) and the Fundamental
Research Funds for the Central Universities of China.

\vspace{3ex}
\vspace{3ex}

\noindent\textbf{Author contributions}

\noindent Y.P.Li synthesized single crystals and performed the transport measurements
with J.Ma. G.Yao, N.N.Liu, C.H.Liu and J.F.Jia carried out the STM measurements.
Y.Wu, Y.Liu, J.Z.Ma, M. Shi and Y. B. Huang performed the ARPES measurements.
C.C.Xu and C.Cao did the DFT calculations. Y. Liu helped do samples
using a focused-ion-beam system to observe a atomic phase by HRTEM. B.Hua, X.H.Yang, L.Qiao and M.C.Li
helped grow the samples. B.J.Lv and H.Xing performed the transport measurement at extremely low temperature.
Y.P.Li, Y. Liu and Z.A.Xu wrote the manuscript.
All the authors contributed to the discussion of results
and improvement of the manuscript.

\vspace{3ex}

\noindent\textbf{Additional information}

\noindent\textbf{Competing interests:} The authors declare no competing interests.







\clearpage
\newpage

\onecolumngrid
\renewcommand{\figurename}{FIG.S}
\setcounter{figure}{0}
\setcounter{table}{0}
\setcounter{equation}{0}
\textbf{Supplementary Materials}

\begin{figure*}[!thb]
\begin{center}
\includegraphics[width=5in]{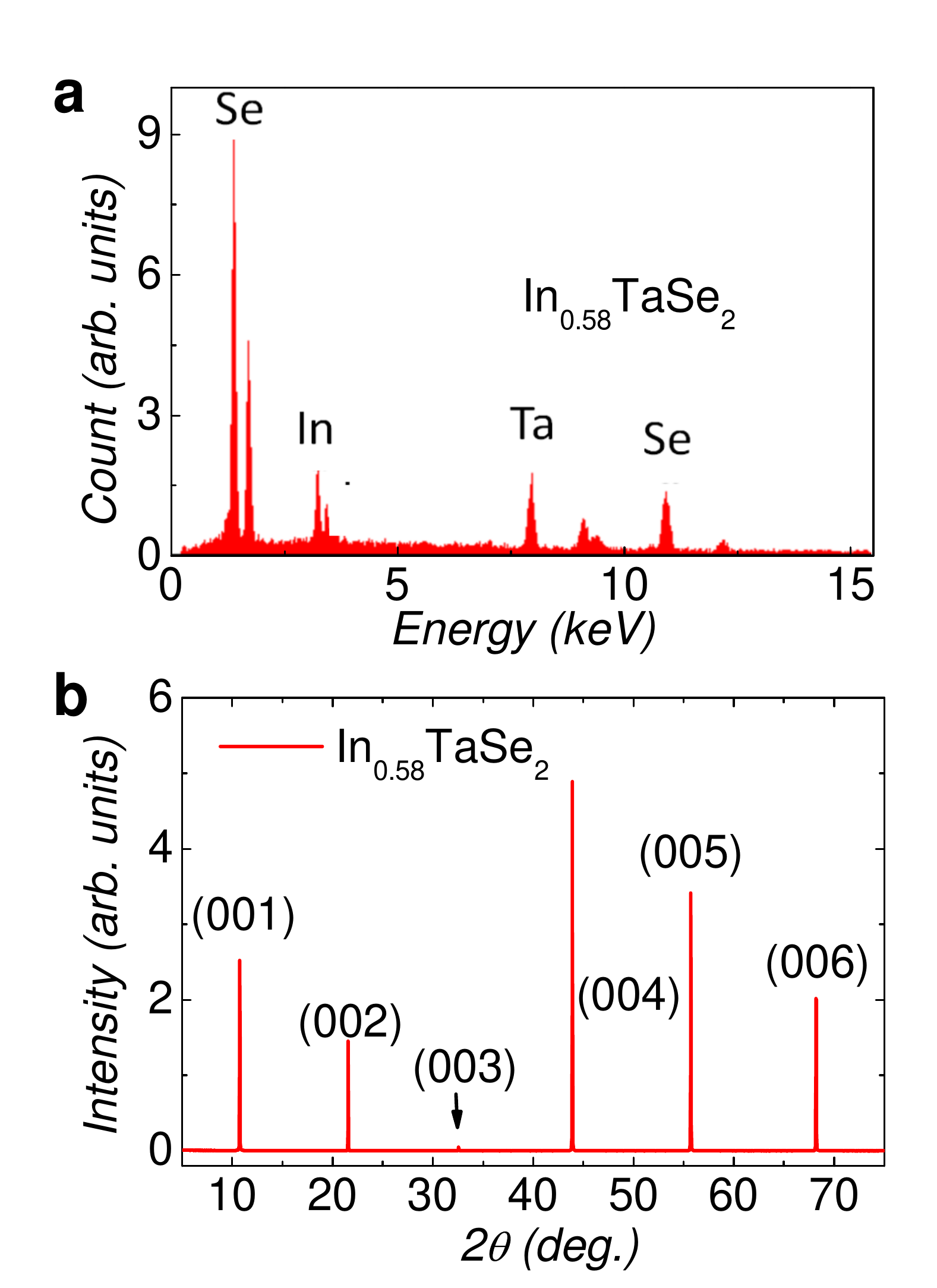}
\end{center}
\caption{\label{Fig.S1} \textbf{EDS and XRD data of In$_{0.58}$TaSe$_{2}$.} \textbf{a}, The EDS
spectrum of In$_{x}$TaSe$_{2}$ (S0 sample) and resultant chemical composition In:Ta:Se = 0.58:1:2.06. \textbf{b}, The single crystal XRD pattern for the (001) facet of In$_{0.58}$TaSe$_{2}$ (S0 sample).     }
\end{figure*}

\begin{figure*}[!thb]
\begin{center}
\includegraphics[width=6.5in]{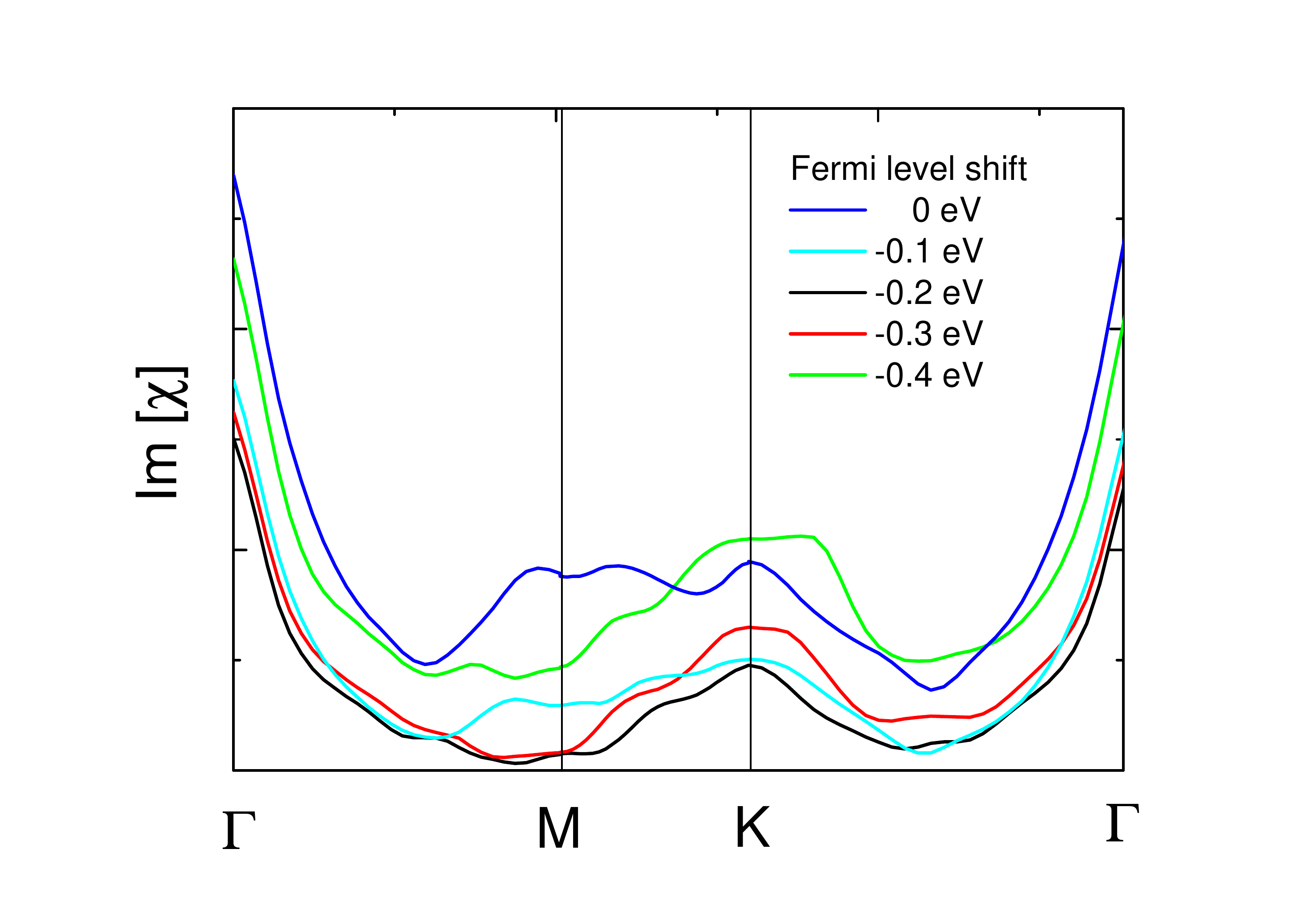}
\end{center}
\caption{\label{Fig.S2} \textbf{The imaginary parts of susceptibility $\lim\limits_{\omega \to 0}$ Im$[\chi_{0}(\omega,\textbf{q})]/\omega$ of InTaSe$_{2}$ calculated with different Fermi energy shifts.} These imaginary parts reflect the Fermi surface nesting. At the 0 eV shift values (blue line), a very weak peak emerging at around $M$, which should correspond to the 2 $\times$ 2 CDW wave vector observed experimentally. We also note that the imaginary part of susceptibility is also highly sensitive to the Fermi level shift, and hence the indium content $x$. Therefore, other mechanisms could also be relevant for the CDW formation.        }
\end{figure*}

\begin{figure*}[!thb]
\begin{center}
\includegraphics[width=4.0in]{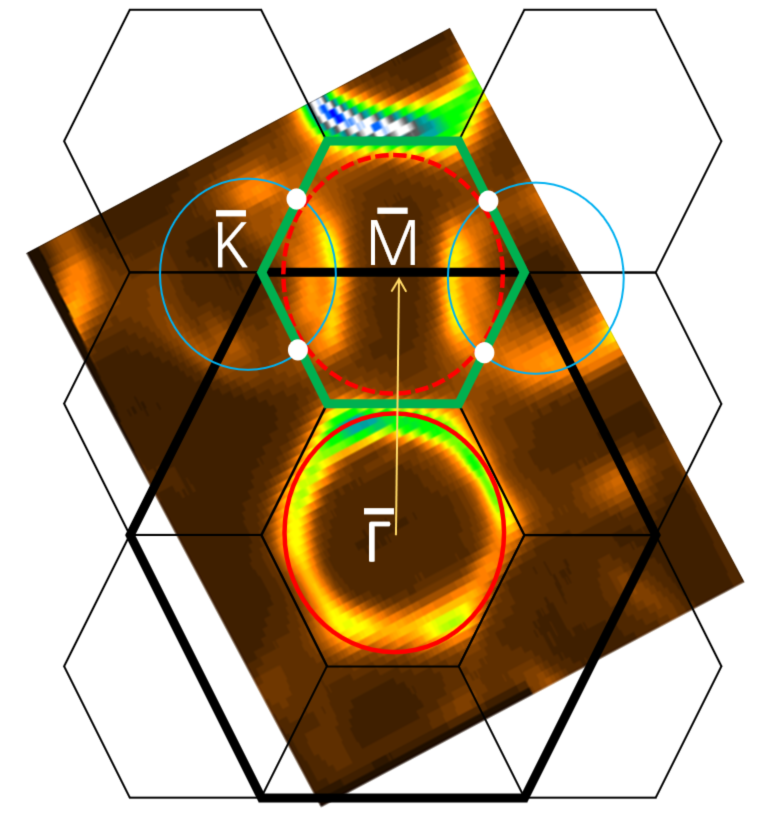}
\end{center}
\caption{\label{Fig.S3} \textbf{A schematic diagram showing the possible mechanism of anisotropic CDW gap opening in In$_{x}$TaSe$_{2}$.} The thin and thick black hexagons are the Brillouin zones for 1 $\times$ 1 and 2 $\times$ 2 superlattices, respectively. The red and blue ellipses represent the hole pocket and electron pockets, respectively. The dashed red ellipse is the CDW folded replica of the original bulk hole pocket at $\bar{\Gamma}$ point (red ellipse). The CDW ordering vector is represented by the orange arrow. The white dots highlight the crossing points of the main bands and the (weak) CDW folded bands, where the CDW gaps are observed experimentally.          }
\end{figure*}

\begin{figure*}[!hb]
\begin{center}
\includegraphics[width=6.5in]{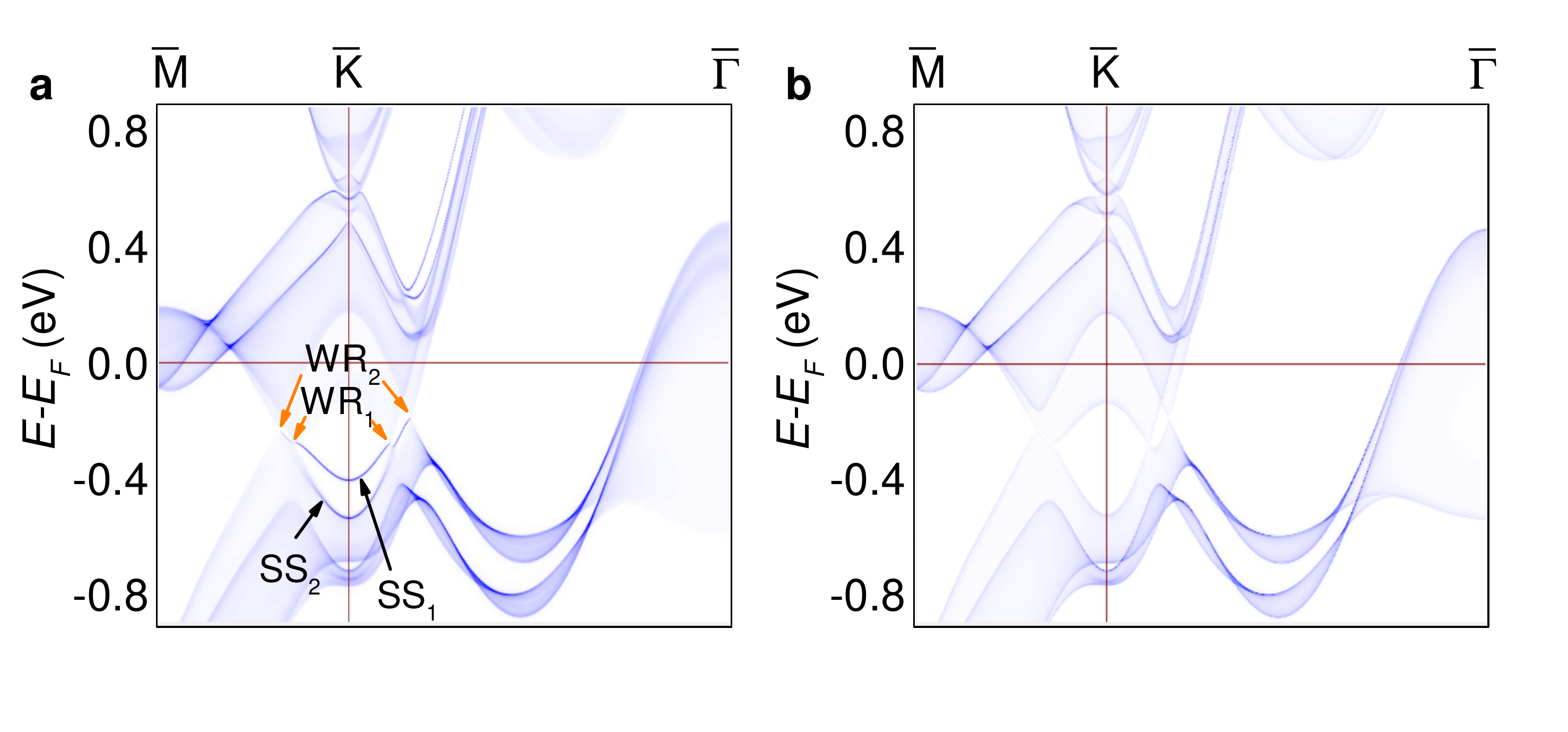}
\end{center}
\caption{\label{Fig.S4} \textbf{Surface states in InTaSe$_{2}$ from the
DFT calculations.} \textbf{a}, The band
structures are projected on the (001) facet. \textbf{b}, Only bulk states. Compared with the
bulk bands only in \textbf{b}, the surface states are marked by SS$_{1}$
and SS$_{2}$ in \textbf{a}. SS$_{2}$ is merged into the bulk continuum
and therefore is difficult to identify experimentally.}
\end{figure*}

\begin{figure*}[!thb]
\begin{center}
\includegraphics[width=6in]{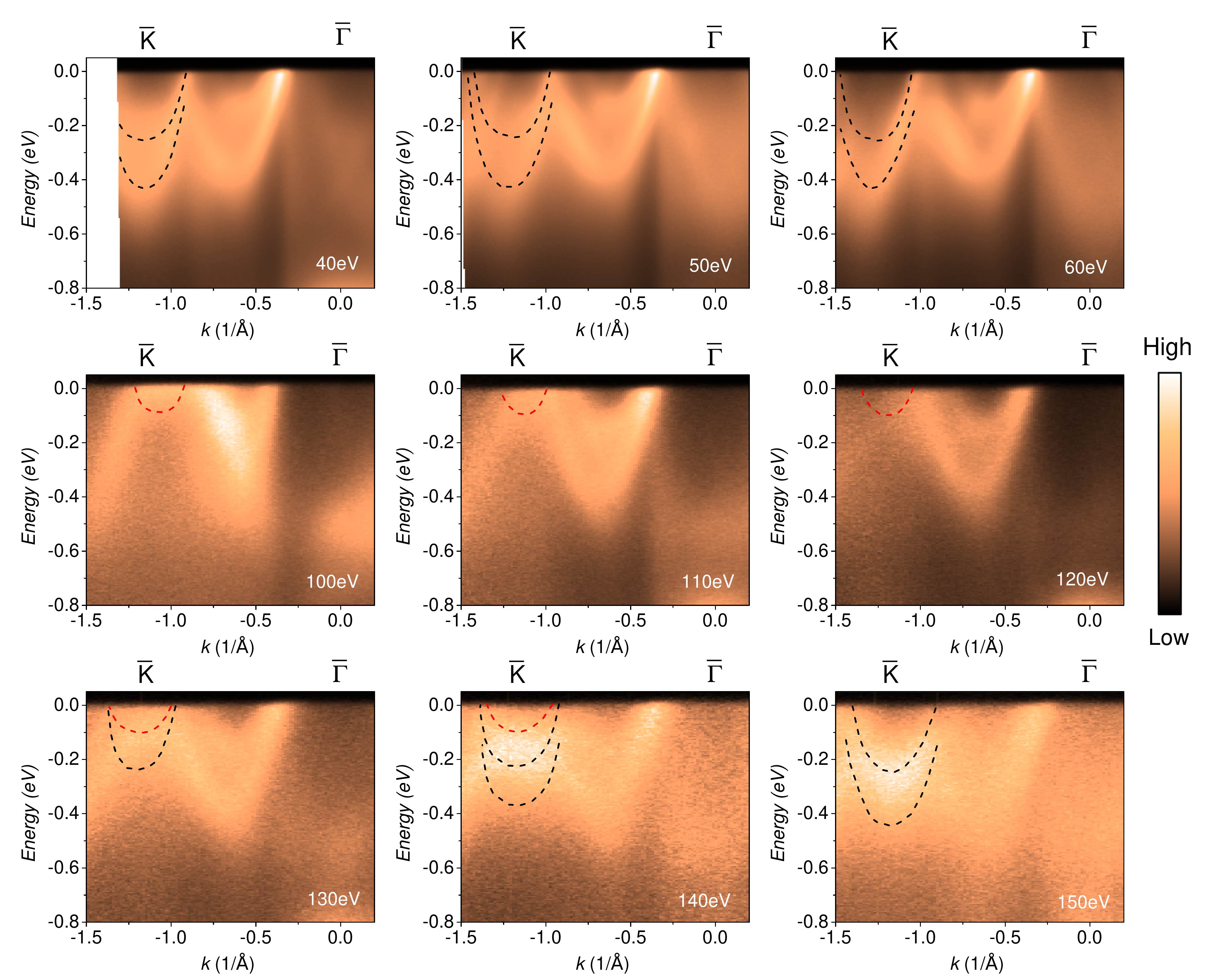}
\end{center}
\caption{\label{Fig.S2} \textbf{Large-range photon energy dependent ARPES data.} The black dashed lines indicate In 5p bulk bands, which do not show large variation with the photon energy likely due to weak interlayer coupling, similar to the observations in PbTaSe$_{2}$ \cite{PbTaSe2_BianG_NatC}. The red dashed line near the Fermi level indicates the possible surface state SS$_{1}$, which could only be observed in a small photon energy window from 100 eV to 140 eV.        }
\end{figure*}

\begin{figure*}[!thb]
\begin{center}
\includegraphics[width=6.5in]{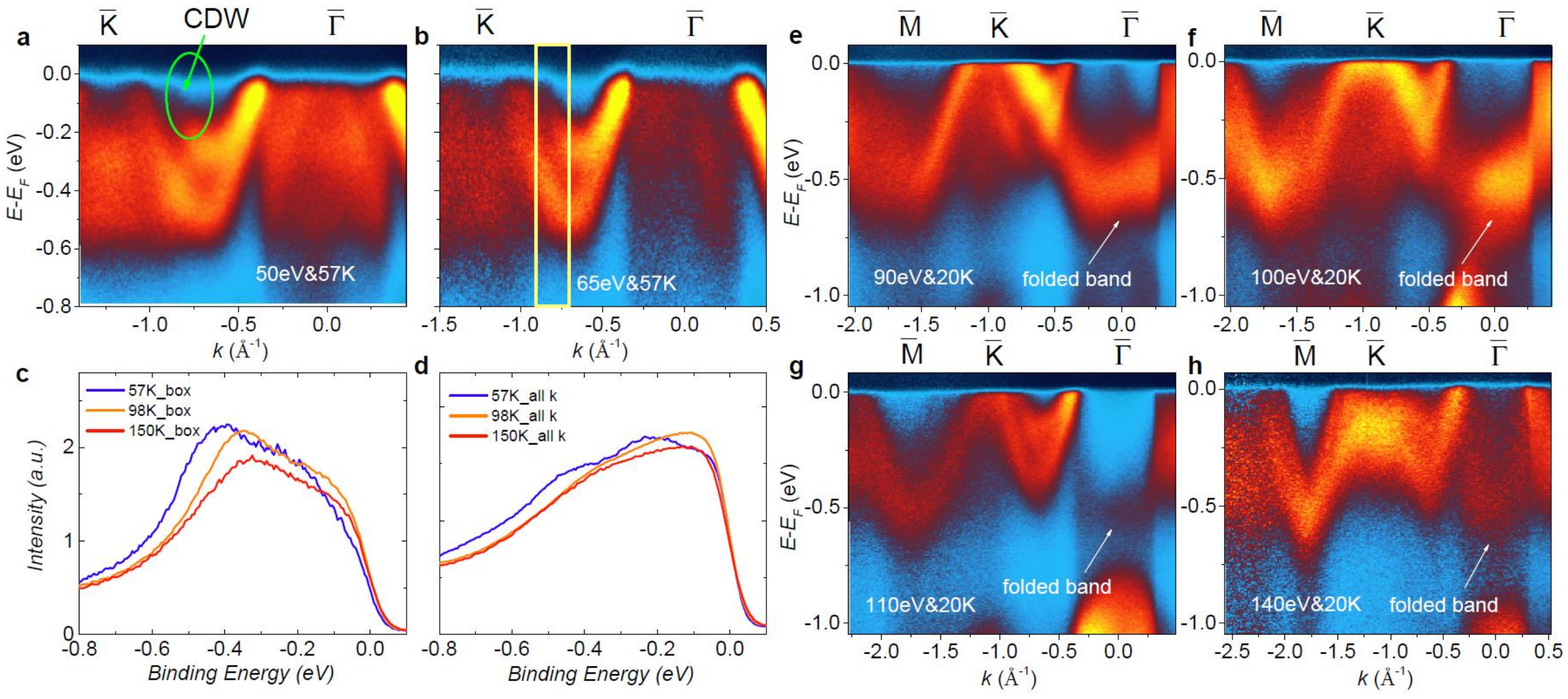}
\end{center}
\caption{\label{Fig.S4} \textbf{Additional ARPES data for In$_{x}$TaSe$_{2}$.} Energy-momentum cuts of another In$_{x}$TaSe$_{2}$ sample taken with 50 eV (\textbf{a}) and 65 eV photons (\textbf{b}) at 57 K, showing similar gapping of the outer Weyl ring along $\bar{\Gamma}$ - $\bar{K}$ direction as in Fig. 3d. The CDW gap is marked by a green circle. \textbf{c}, Energy distribution curves (EDC) integrated within the yellow box in \textbf{b} show suppression of spectral intensity at E$_{F}$ at low temperature (i.e., opening of the CDW gap) and simultaneous downward shift of the valence bands. Note that the downward shift seems to be consistent with the reduction of hole carriers at low temperature from Hall measurements shown in Fig. 2e, although its origin is still unclear at the moment. \textbf{d}, Integrated EDCs for all momenta in \textbf{b}, showing similar downward shift of the overall valence bands and not a fully opened gap at the Fermi level (due to dominant contributions from other ungapped FS regions). \textbf{e},\textbf{f},\textbf{g},\textbf{h}, CDW folded bands can be detected at $\bar{\Gamma}$ (white arrows), as a result of 2$\times$2 CDW folding from the main bands at $\bar{M}$ point.    }
\end{figure*}

\begin{figure*}[!thb]
\begin{center}
\includegraphics[width=6.5in]{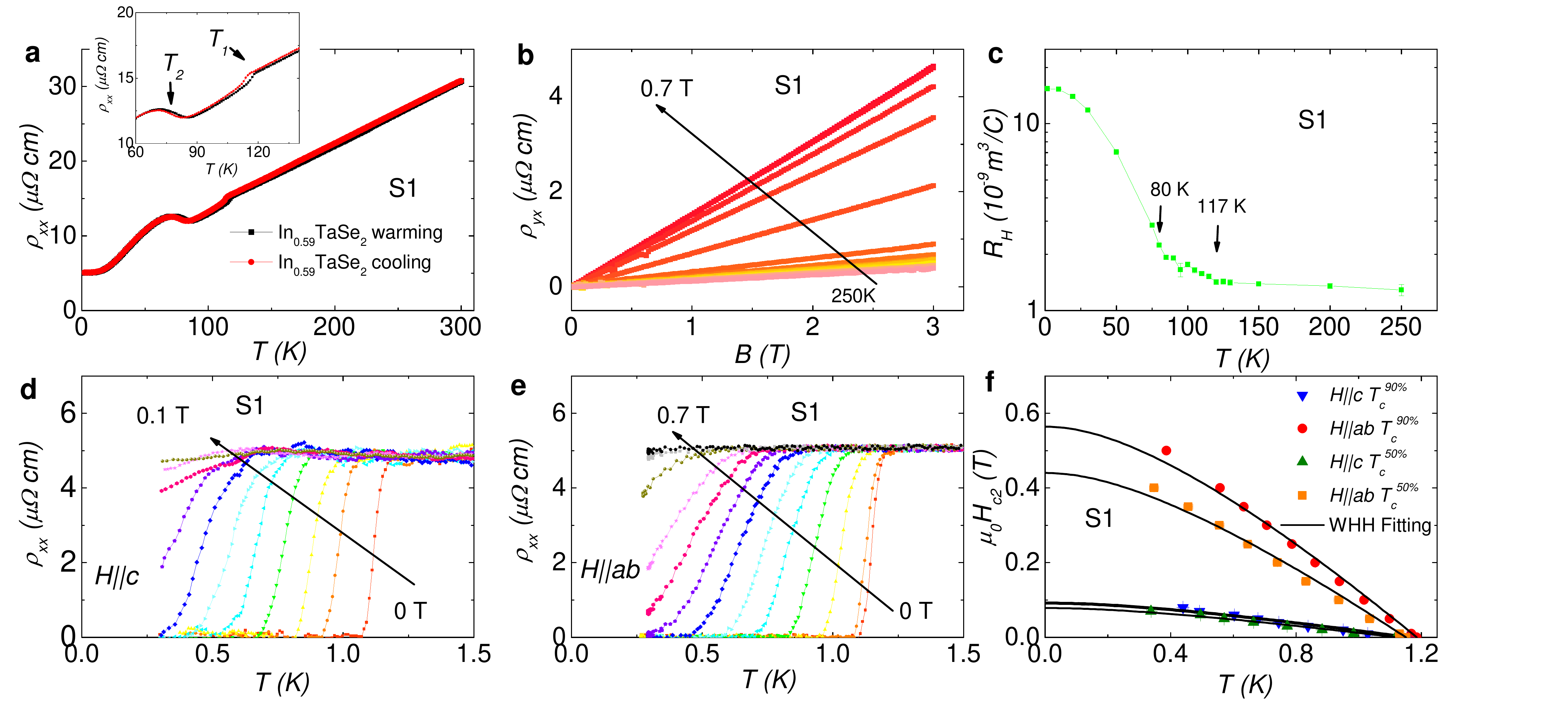}
\end{center}
\caption{\label{Fig.S5}\textbf{Characterization of the sample S1.} \textbf{a}, $\rho_{xx}$ of sample S1 for temperature warming and cooling. The inset is the enlarged view, indicating very weak hysteresis at both CCDW transitions.
\textbf{b}, Magnetic-field dependent $\rho_{yx}$ at different temperatures. \textbf{c}, Hall coefficient increases distinctly below 120 K. \textbf{d},\textbf{e}, The superconducting transitions are obtained under various magnetic fields
when $H\|c$ and $H\|ab$, respectively. \textbf{f}, Magnetic-field-dependent $T_{c}^{90\%}$ and $T_{c}^{50\%}$ are plotted for $H\|ab$ and $H\|c$. The upper-critical-fields are fitted by the WHH model, and the $H_{c2}$ at low temperature deviates the fitting, which suggests the possibility of unconventional superconductivity in In$_{x}$TaSe$_{2}$.     }
\end{figure*}

\begin{figure*}[!thb]
\begin{center}
\includegraphics[width=6in]{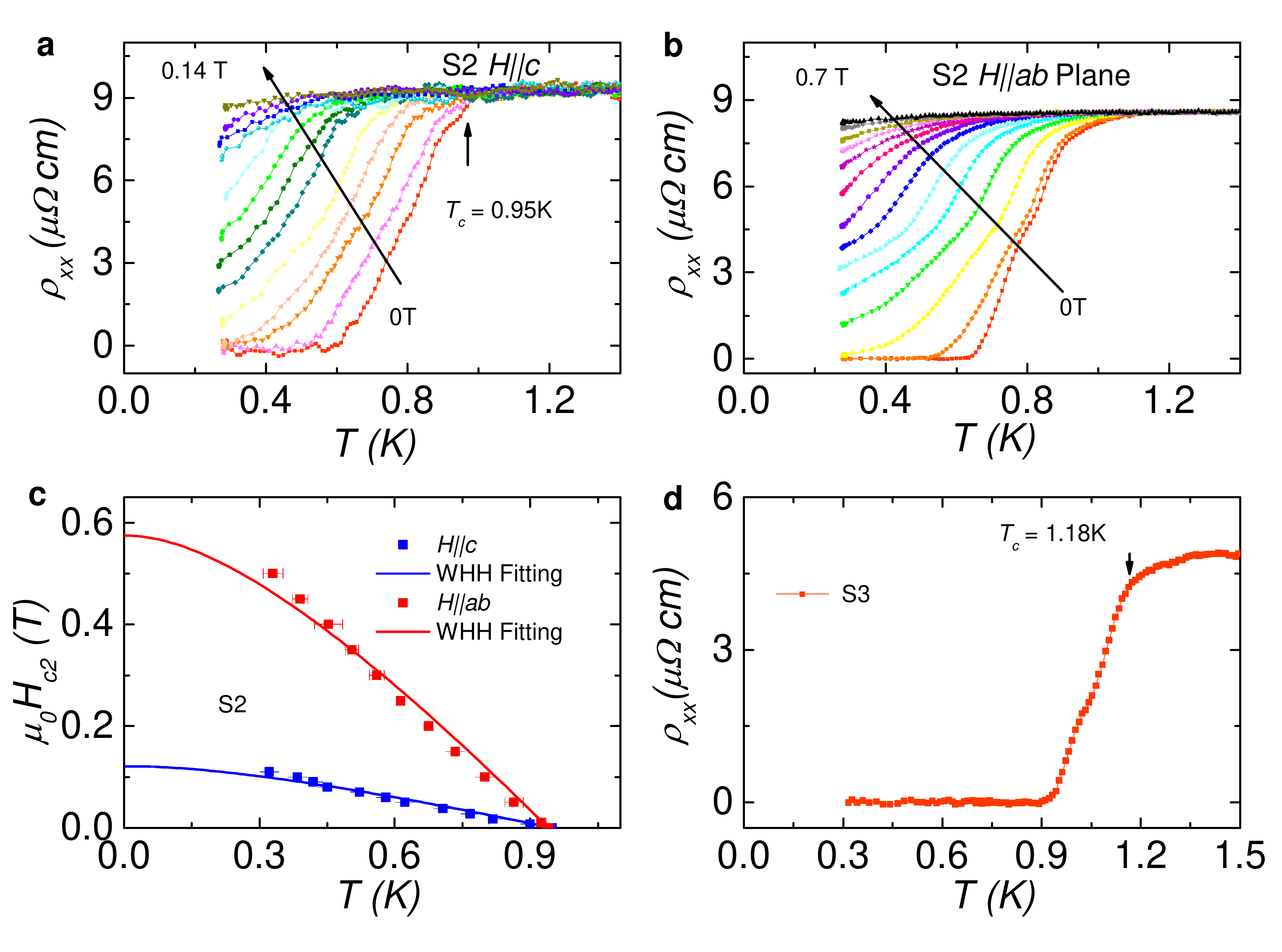}
\end{center}
\caption{\label{Fig.S6}\textbf{Superconductivity in other In$_{x}$TaSe$_{2}$ samples with $x \sim 0.6$.}
Superconductivity at $H\|c$ (\textbf{a}) or $H\|ab$ (\textbf{b}) is shown in the sample S2 with
$T_{c}$ = 0.95 K. \textbf{c}, Phase diagram of $H_{c2}$ vs. $T$ for $H\|ab$ and $H\|c$. The anisotropy factor $H^{H\|ab}_{c2}$/$H^{H\|c}_{c2}$ is 4.7 in the sample S2.
\textbf{d}, Another sample (S3) also displays the superconducting transition with $T_{c}$ = 1.18 K. More samples are summarized in the Table.\ref{SC}.     }
\end{figure*}

\clearpage

\begin{table}[!thb]
\tabcolsep 0pt \caption{\label{SC}\textbf{Summary of several In$_{x}$TaSe$_{2}$ samples.} These samples grown by chemical vapor transport method almost have the same In content with $x \sim 0.6$, and the transition temperatures $T_{c}$, $T_{1}$ and $T_{2}$ vary slightly. Transport properties shown in the main text are measured on the sample S0.   } \vspace*{-12pt}
\begin{center}
\def\temptablewidth{1.0\columnwidth}
{\rule{\temptablewidth}{1pt}}
\begin{tabular*}{\temptablewidth}{@{\extracolsep{\fill}}cccccc}
Sample  &$x$    &$T_{1}$    &$T_{2}$   &$T_{c}$  &$H^{H\parallel ab}_{c2}$/$H^{H\parallel c}_{c2}$(0 T, $T_{c}^{90\%}$)   \\ \hline
S0      &0.58     &116 K         &77 K    &0.91 K    &4.6  \\
S1      &0.59     &117 K         &80 K    &1.18 K    &6.1   \\
S2      &0.58     &114 K         &78 K    &0.95 K    &4.7  \\
S3      &0.57     &118 K         &83 K    &1.18 K    &-- \\
S4      &0.57     &117 K         &80 K    &1.36 K    &-- \\
S5      &0.59     &116 K         &75 K    &1.45 K    &--  \\
S6      &0.59     &113 K         &76 K    &1.45 K   &-- \\
S7      &0.6      &116 K         &77 K    &0.84 K    &-- \\
\end{tabular*}
{\rule{\temptablewidth}{1pt}}
\vspace*{-18pt}
\end{center}
\end{table}

\clearpage

\begin{thebibliography}{10}
\bibitem{CDWInTaSe2}
\bibinfo{title}{{The detail CDW transiion will be discussed in another work}} .
\end{thebibliography}

\end{document}